# Optimization of Inverted Double-Element Airfoil in Ground Effect using Improved HHO and Kriging Surrogate Model

Paras Singh[1*], Arun Ravindranath[1], Aryan Tyagi[1],
Aryaman Rao[2], Raj Kumar Singh[1]

[1] Department of Mechanical Engineering, Delhi Technological University, New Delhi, India

[2] Department of Electrical Engineering, Delhi Technological University, New Delhi, India

[*] parass2802@gmail.com

## ABSTRACT

In the automotive industry, multi-element wings have been used to improve the aerodynamics of race cars. Multi-element wings can enhance a vehicle's handling and stability by reducing drag and increasing downforce, allowing it to corner more effectively and achieve higher speeds. Performance gains by utilizing the ground effect are highly sensitive to the wing setup. This study focuses on identifying the optimum design parameters for the airfoil to achieve the desired downforce and drag performance. The design parameters chosen are ride height, flap overlap, flap angle, and flap gap (the spacing between the flap and the main airfoil). These parameters are optimized for three different use cases: high downforce, low drag, and a setup with the highest airfoil efficiency. The force coefficient and flow field data were gathered using two-dimensional (2D) Reynolds Averaged Navier Stokes (RANS) simulations, with the turbulent flow modeled using the k-ω Shear Stress Transport (SST) turbulence model. The Improved Harris Hawks Optimization (HHO) algorithm was used to obtain the optimal configuration of the double-element and the resulting designs showed a significant improvement in downforce and drag performance compared to the baseline designs. Improved HHO was further compared with other state-of-the-art algorithms for assessing the algorithm's performance for a problem with highly non-linear behavior, where it was able to demonstrate its ability to obtain the optimal solutions more efficiently.

**Keywords:** Computational Fluid Dynamics; Aerodynamics; Multi-Element Airfoils; Ground Effect; Harris Hawks Optimization; Surrogate Model

# 1 Introduction

Historically, aerodynamics typically played a significant role in aircraft development, where it was an essential aspect of the design process. In contrast, it did not usually hold as much importance in the development of road vehicles, where factors like aesthetics often took precedence. However, racing cars were an exception; they strategically utilized airflow to enhance cooling and performance [1]. In general, a car's aerodynamic performance can be enhanced in two ways: by reducing drag, which leads to higher straight-line speeds, or by increasing negative lift (downforce), which improves the car's grip for faster cornering. Improving downforce often increases drag, and vice versa, necessitating a balance between the two to optimize performance based on the specific track layout [2].

The majority of aerodynamic development on open-wheel racing cars has been on improving downforce, thereby improving performance in zones with limited traction such as high-speed corners. This led to inverted airfoil wings having found use in improving the aerodynamics of race cars, the renowned Lotus 49B being one of the first cars to showcase inverted wings in 1967 [3]. By reducing drag and increasing downforce, inverted wings were found to enhance a vehicle's handling and stability, allowing it to corner more effectively and achieve higher speeds. By 1984, McLaren developed multi-element wings for the MP4/2 which led to inverted multi-element wings in ground effect being used within the automotive industry extensively in the years that followed [4].

When positioned near a ground plane, an inverted wing creates a more intricate flow pattern in contrast to the free-stream scenario. In the case of multi-element wings in free-stream conditions, the resulting flow pattern is predominantly influenced by viscous effects and includes areas with streamlines with pronounced curvature, the merging of various boundary layers, and the convergence of wake regions [5]. Hence, the interaction between a multi-element airfoil's flow field and the ground effect conditions results in a significantly intricate, mutually influencing flow pattern primarily characterized by viscous effects.

The study of multi-element flows has presented a longstanding challenge to researchers, both in experimental and computational domains. In 1978, Smith [6] presented a proof stating that double-element airfoils are better than single-element airfoils, thus by the nature of the development, proving that multi-element airfoils are better than lone airfoils in certain conditions. Twenty years later, Ranzenbach et al. [7] conducted investigations on an inverted airfoil (NACA $63_2$-215 Mod B) using computational and experimental methods and compared the results with road conditions. They carried out Reynolds-averaged Navier-Stokes or RANS simulations while considering a moving ground at a Reynolds number (Re) of $1.5 \times 10^6$ based on the airfoil's chord length. The grid used consisted of only $3 \times 10^4$ cells, considerably coarser than current investigations which usually use grids in the magnitude of $6 \times 10^5$ cells.

In the years following Ranzenbach, Zerihan [8], Zerihan and Zhang [9], and Zhang and Zerihan [10-12] conducted various experiments on a 1:1.25 scale model of the Tyrrell 026 car's front wing, which competed in the 1998 Formula One championship. These experiments have arguably become the primary reference for researchers seeking validation in the study of single and multi-element wings operating in ground effect. The evaluations were conducted using laser Doppler anemometry and particle image velocimetry, along with the collection of force balance data to assess both drag and downforce. Mahon and Zhang [13,14] used these results to conduct computational investigations into a double-element airfoil in ground effect and concluded that the main element of the airfoil was responsible for a major part of the downforce generated while the flap contributed towards most of the drag generated by the wing. It was also deduced that the overall thickness of the wake created by the wing increased with a reduction in ride height due to the spreading rate of the wake beneath the main element vastly increasing as the ride height was reduced.

Conventionally, design engineers relied manually on design processes with the help of their intellect and expertise. However, over the past years, the complexity of these design processes has increased attributable to intricate factors such as design constraints, multi-body dynamics, and non-linearity. This inconvenience led to the adoption of classical optimization techniques to accurately estimate the optimal parameters employing the gradients of various objective functions in the field of design mechanics. Although the classical methods had their perks, they suffered from getting trapped in the local optima of the objective and their time complexity was quite high, making them unfeasible for real-world applications. The introduction of metaheuristics optimization gained immense popularity in the domain of mechanical engineering spanning from aerospace or aircraft design [17] to automotive industries [18]. Metaheuristics overcame premature convergence without the assistance of objective function gradients which can be time-consuming. This popularity prompted us to incorporate Improved HHO to acquire the optimal parametric configuration for the double-element wing for improved drag and downforce performance.

Despite these experiments, there still remain gaps in knowledge regarding the optimal performance of multi-element inverted airfoils in ground effect. The intricate aerodynamic interactions, the requirement to optimize ground clearance, the difficulties with computational modeling, the difficulties of design optimization, and the requirement for experimental validation all contribute to the complexity of double-element airfoil optimization in the case of ground effect. To increase the setup's overall effectiveness, it becomes necessary to optimize the various characteristics of the multi-element wing for diverse usage scenarios. This study focuses on computationally identifying the optimum design parameters for the double-element airfoil to achieve the desired downforce and drag performance.

## 2    Methodology

### 2.1   Airfoil Layout and Parameterization

The optimization parameters chosen are ride height, flap overlap, flap angle, and flap gap. These parameters are optimized for three different use cases: high downforce, low drag, and a setup with the highest airfoil efficiency.  This study uses the data provided by Zhang and Zerihan [10]. The model used is the front wing of the racecar Tyrrell Racing built for the 1998 Formula One Championship (Tyrrell 026) at a 1:1.25 scale, which in turn used a modified version of the airfoil created by NASA, here referred to as the GA(W) LS (1)-0413 airfoil. This model was employed due to the extensiveness of the experiments, along with data for various aerodynamic forces like downforce and drag, and pressure measurements along the span and chord for a variety of angles of attack and ride heights [17]. For the present study, the angle of incidence of the main element is held constant at $\alpha_m = 1°$. The ride height h lies within the range of 20 mm to 200 mm and the flap angle $\alpha_f$ lies between 0 to 25°. The range for flap overlap $\delta_o$ is taken to be 30 mm from both sides of the tip of the main element while the range of 9 mm to 40 mm is considered as the flap gap $\delta_g$ range. The schematic of the airfoil configuration is shown in Fig.1 (a).

For modeling the flow physics of the problem accurately a high-quality unstructured tri mesh was used with progressively finer cells near the airfoil and the moving ground. To replicate the actual conditions inside the wind tunnel, a moving ground was placed beneath the airfoil with inflation layers to model boundary layer effects accurately. The initial cell spacing, parallel to the wall, inside the ground and airfoil boundary layer blocks, was determined to guarantee that $y^+ \approx 1$. The ground and airfoil surfaces were represented as solid walls with a no-slip constraint. The translational velocity of the ground surface was set equal to the freestream to accurately represent the ground effect. The mesh around the airfoils and the moving ground is presented in Fig.1 (b).

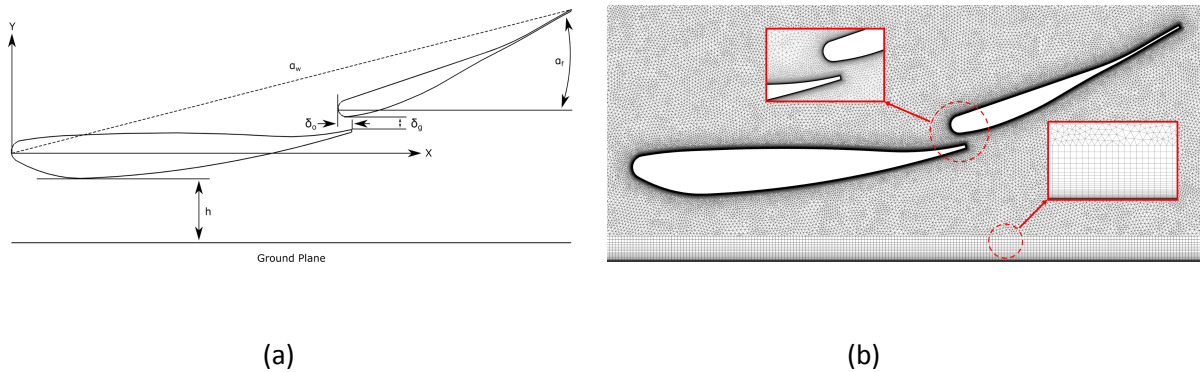

(a)　　　　　　　　　　　　　　　　　　(b)

Fig 1: (a) Definition of terms and coordinate system, (b) Mesh near the airfoils and the moving ground

## 2.2 Setup and Boundary Conditions

Steady-state RANS simulations were conducted using the k-ω Shear Stress Transport (SST) turbulence model for closure. The inlet was set to a freestream velocity of 30 m/s, with a Reynolds number based on the airfoil's chord length held at $7.86 \times 10^5$. The airfoil was subjected to a no-slip boundary condition, and a pressure outlet condition was applied at the exit. The operational pressure was defined as 101325 Pa, with both inlet and outlet set to a gauge pressure of 0 Pa. Given that the Mach number for the simulation was well below 0.3, an incompressible pressure-based solver was employed. The fluid density (ρ) was kept constant at 1.225 kg/m^3 and dynamic viscosity (μ) was maintained at

$1.7894 \times 10^{-5}$ kg/m.s. Non-dimensionalized force coefficients ($C_l$ and $C_d$) were derived using a reference area, A = 1.6 m^2, and a reference length, l = 1 m. The pressure-velocity coupling utilized the COUPLED algorithm, and gradient discretization was carried out using a least square cell-based scheme. Spatial discretization for momentum, pressure, specific dissipation rate, and turbulent kinetic energy employed a second-order linear upwind scheme. To ensure convergence, the simulation underwent 1000 iterations for all cases, until $C_l$ and $C_d$ monitors, along with the residuals, reached the predetermined limit of $1 \times 10^{-5}$.

## 2.3 Grid Independence Test and Model Validation

To examine the sensitivity of the numerical solution on the grid size, a grid independence test was conducted. For the test, three different grids were constructed, namely Coarse, Medium, and Fine, with their cell counts being $2 \times 10^5$, $4 \times 10^5$, and $8 \times 10^5$ respectively. In addition, the numerical results were validated against experimental data [4]. The results of the study are summarized in Fig.2. It can be observed that both the fine and medium grids work well at modeling the behavior of $C_l$ and $C_d$ for the double-element airfoil. Also, the deviation between the values of $C_l$ and $C_d$ for the fine and medium grids is quite small and thus we can infer that grid convergence has been achieved. Hence, to save computational resources, the medium grid will be utilized for further parts of the study.

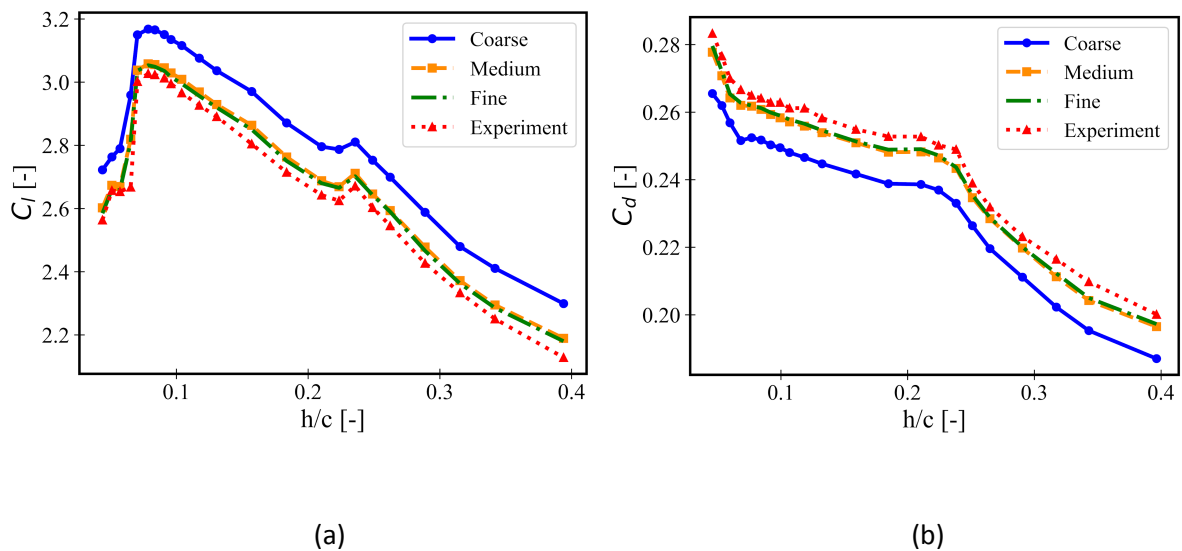

(a)          (b)

Fig 2: Comparison of (a) $C_l$ and (b) $C_d$ for the double-element airfoil computed using three different grid resolutions (Coarse, Medium, and Fine) with experimental data [4]

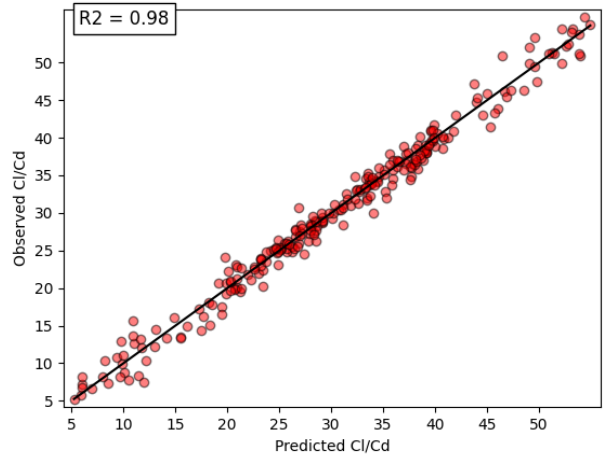

Fig 3: Validation of Kriging Surrogate Model

## 2.4 Optimization Methods

The optimization methodology involved the integration of the Kriging surrogate model with the Improved Harris Hawks Optimization algorithm [7]. To construct the surrogate model, a training dataset was obtained through RANS simulations of the double-element airfoil. Sobol sampling method was employed to generate 200 samples for the training data. The independent variables, as discussed above, were ride height, flap overlap, flap angle, and gap. The objective function was chosen appropriately for each setup: minimization of drag coefficient for the low drag setup, maximization of lift coefficient for high downforce setup, and maximization of $C_l/C_d$ for maximum airfoil efficiency setup. The validation of the Kriging surrogate model performed using 200 additional unseen data points is shown in Fig.3.

The trained surrogate model served as a valuable tool to approximate the complex aerodynamic behavior of the airfoil, facilitating efficient exploration of the design space. To determine the optimal set of parameters, the Improved Harris Hawks Optimization algorithm was employed. This evolutionary algorithm, inspired by the hunting strategies of Harris's Hawks, leveraged a combination of exploration and exploitation techniques to iteratively search for the global optimum. The Improved HHO oscillates between two phases: exploration and exploitation phase. In the exploration mechanism, the hawks rely on their keen eyesight to spot and follow potential prey. Yet, there are instances where their prey remains hidden from view. During these times, the hawks exhibit patience, diligently monitoring vacant areas to locate their quarry. Once the prey is spotted, they transition into their hunting phase, strategizing their descent for the capture. Typically, the prey, often a rabbit, represents the optimal candidate solution, while the hawks themselves symbolize the other potential solutions.

$$E = 2E_0(1 - \frac{t}{T}) \tag{1}$$

$$X(t+1) = X_{rand}(t) - r_1|X_{rand}(t) - 2r_2 X(t)|, \text{ if } q \geq 0.5 \tag{2}$$

Depending on the escaping energy E parameter as shown in Equation 1, the hawks effortlessly shift from exploration to exploitation phase. $E_0$ indicates the initial energy state in the equation. 1. The exploration focuses on following the prey until it has run out of its escaping energy and then finally encircling and attacking the prey, while the exploitation aims at ambushing the prey with a surprise pounce once it is exhausted. The equations (2-3) represent the updation procedure for hawks in the exploration phase and similarly, the hawks attack their prey or achieve ideal solutions through different modes in the exploitation phase. In the equation, X(t) represents the solution at time t, $X_{rabbit}$ denotes the best or optimal solution and $X_m$ (Eq. 4) signifies the mean solution of hawks. The parameters $r_1$, $r_2$, $r_3$, $r_4$ denote random values between 0 and 1, LB and UB represent lower and upper bounds of the optimization function. The Improved HHO effectively addresses the issue that traditional HHO often encounters which is becoming trapped in local optima when dealing with complex objective functions. This issue is resolved by incorporating functions such as a Levy-Flight, which enables the hawks to thoroughly navigate the search space, and effectively find optimal solutions.

$$X(t+1) = (X_{rabbit}(t) - X_m(t)) - r_3(LB + r_4(UB - LB)), \text{ if } q < 0.5 \tag{3}$$

$$X_m(t) = \frac{1}{N}\sum_{i=1}^{N} X_i(t) \tag{4}$$

# 3    Results and Discussion

## 3.1    Performance against state-of-the-art optimization algorithms

Table 2 displays the result of Improved HHO as compared to different baseline algorithms such as Bat algorithm **[19]**, Cuckoo Search Algorithm [20] and Grey Wolf Optimisation **[21]**. It can be clearly observed from Figure 4 that Improved HHO outperforms the other algorithms by achieving the best fitness value for all of the three cases.

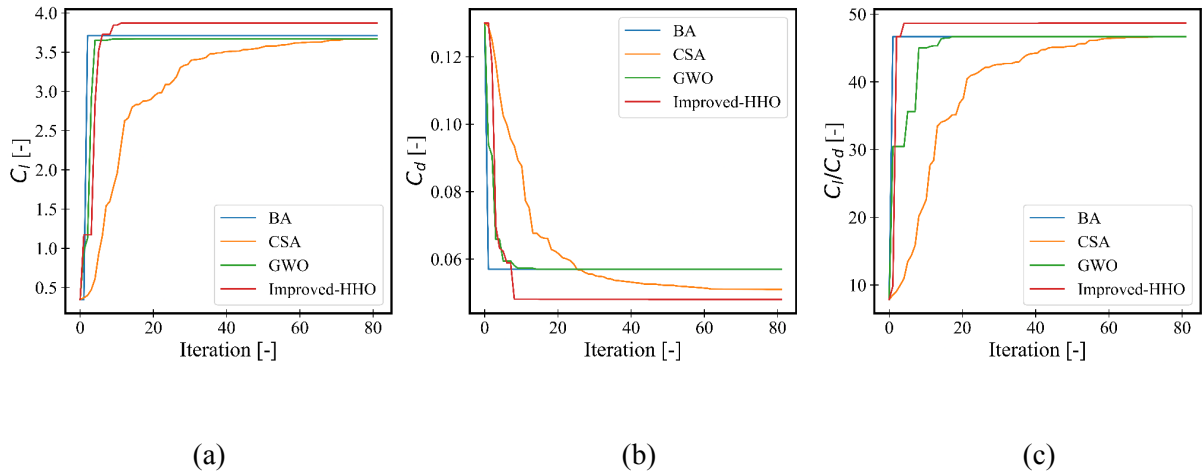

(a)                                          (b)                                          (c)

Fig 4. Comparison of metaheuristic algorithms for (a) maximizing $C_l$ (b) minimizing $C_d$ (c) maximizing $C_l/C_d$

## 3.2    Optimum parameters

The results of the study are summarized in Table 3. Here Setup A (h= 142 mm, $\delta_o$= -2 mm, $\delta_g$= 18 mm, $\alpha_f$= 15.7º) refers to the initialized design point for the optimization algorithm, with the values of objective function being as follows - (a) $C_l$= 2.83, (b) $C_d$= 0.086, (c) $C_l/C_d$= 32.91. Setup B (h=20 mm, $\delta_o$= -12 mm, $\delta_g$= 13.8 mm, $\alpha_f$= 22.5º)  is the high downforce setup, Setup C (h=182 mm, $\delta_o$= 23 mm, $\delta_g$= 33.2 mm, $\alpha_f$= 6.3º) is the low drag setup and Setup D ( h=56 mm, $\delta_o$= 8.7 mm, $\delta_g$= 21.9 mm, $\alpha_f$= 9.5º ) is the setup with highest airfoil efficiency. From the table, it can be observed that the configurations obtained by the Improved HHO algorithm lead to significant improvements in the objective function.

Table 3: Comparison of force coefficients between Setup A, Setup B, Setup C and Setup D

| Case | Objective Function [-] | Change in comparison to Setup A [%] |
|---|---|---|
| **Setup B** | $C_l$= 3.87 | 36.75 |
| **Setup C** | $C_d$= 0.048 | 44.19 |
| **Setup D** | $C_l/C_d$= 48.67 | 47.89 |

### 3.3 Velocity Distribution

For understanding the flow features leading to the increments in performance, the velocity distribution around the double element airfoil is presented in Fig. 5 for Setup A, B, C, and D. For downforce production, the pressure differential must be in the vertically downward direction. From the velocity distribution, the strongest high-velocity region is observed for Setup B followed by Setup D, then Setup A, and at last Setup C. The incoming air is accelerated due to the convexity of the airfoil's suction surface and also due to the change in cross-section formed between the airfoil's suction surface and the moving ground. This cross-section first decreases till the lowest point of the main element's suction surface is reached and then it begins to increase on moving towards the leeward side. Hence, the peak velocity is observed close to the lowest point of the main element's suction surface. Due to this, the downforce production is highest for Setup B followed by Setup D, and then A, and at last Setup C. Furthermore, a stronger wake region is also observed for the same order. The wake region is a region of slow-moving air that pulls the object towards the downstream direction. Due to the large angle attack for Setup B, the air has to traverse a sharper curvature to stay attached throughout the length of the airfoil. However as observed from Fig. 5 the incoming air does not have the required energy to traverse such sharp curvature and due to this, it separates leading to a stronger separation region for Setup B and hence the highest overall drag force. In contrast, due to the low angle of attack for the Setup C, the incoming air is able to easily follow the path along the main element and the flap leading to a much weaker wake and thus contributing to its lowest $C_d$ among all the Setups presented.

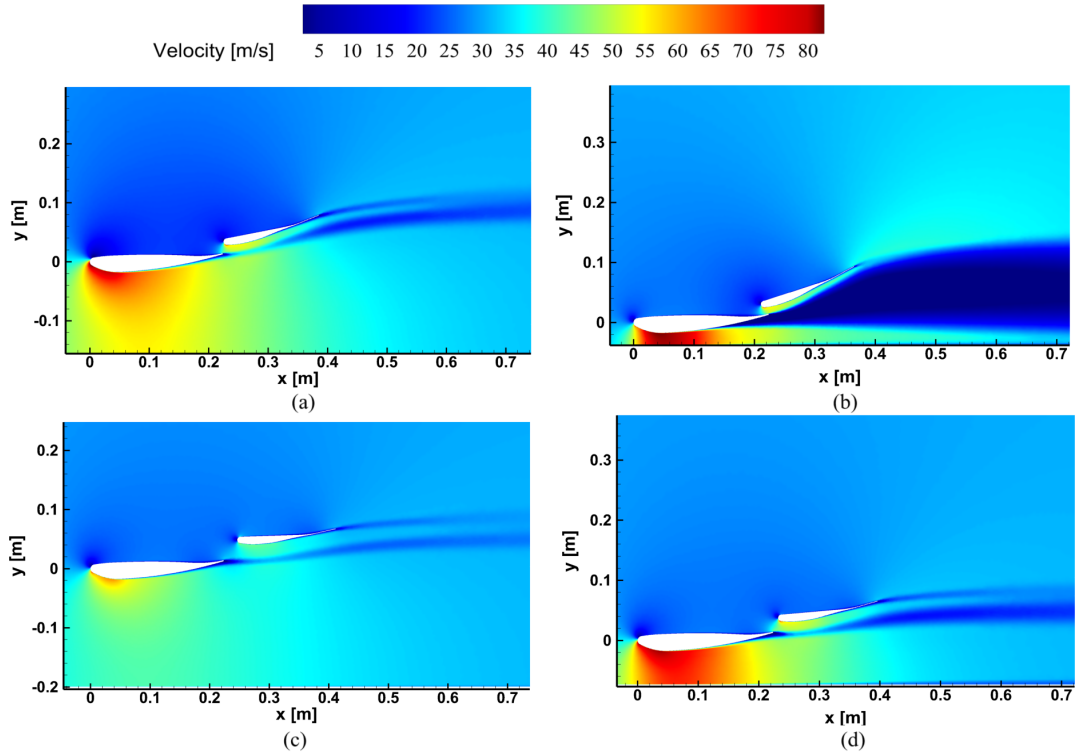

Fig 5: Velocity distribution around (a) Setup A, (b) Setup B, (c) Setup C, and (d) Setup D

### 3.4 Pressure Distribution

The distribution of pressure around the main element and the flap for different setups are presented in Fig. 5. The largest suction pressure is observed for Setup B due to the venturi effect as discussed earlier. This leads to a greater pressure differential between the pressure and the suction side of the double-element airfoil which in turn contributes to the highest negative lift coefficient for Setup B. Setup D has the next higher pressure differential, followed by Setup A and then Setup C. This is also evident from the coefficient of pressure distribution plot shown in Fig. 7 (a). From this plot, it can be inferred that the maximum pressure differential is generated by the main element and thus it is the major contributor towards the overall downforce production which in turn is in line with previous works [11,12,13,14].

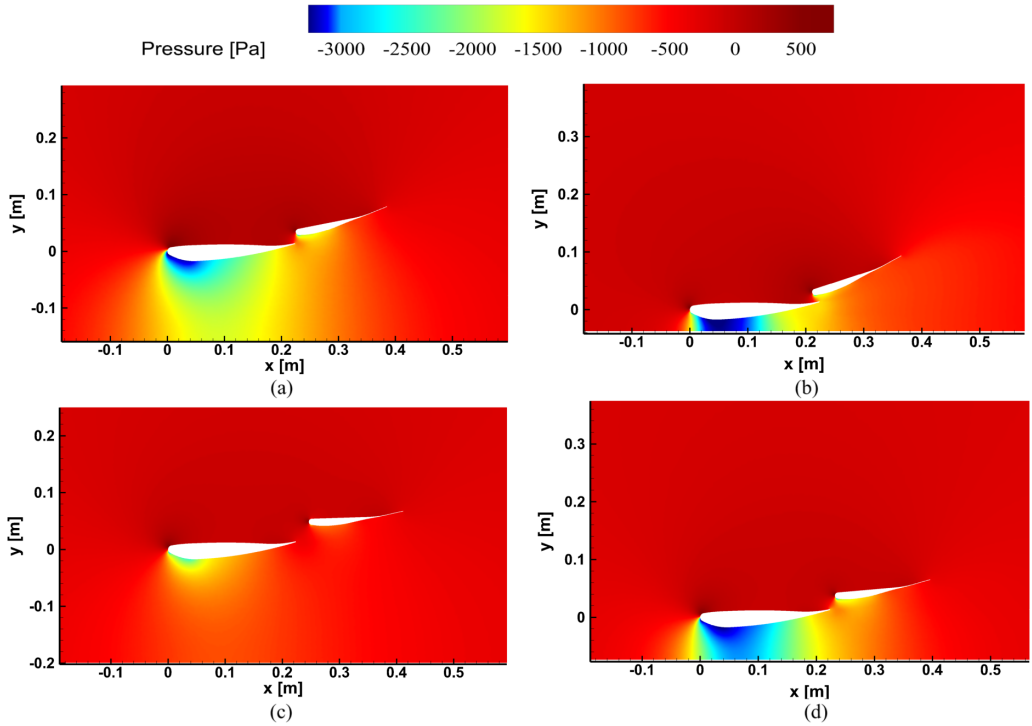

Fig 6: Pressure distribution around (a) Setup A, (b) Setup B, (c) Setup C, and (d) Setup D

## 3.5 Wake velocity Distribution

The wake velocity profiles at a distance of 2%c from the leading edge of the main element in the downstream direction for the different setups are presented in Fig. 7 (b). The strongest upwash effect is created by Setup B, thus leading to its highest downforce generation capability. Furthermore, the highest momentum deficit is there for Setup A which is the initialized design point fed into the optimization algorithm. The lowest momentum deficit in the wake is for Setup C, thus justifying its lowest $C_d$. Finally, Setup D has the second smallest momentum deficit in the wake and the second highest overall pressure differential in the downward direction and also the strength of the up wash effect. Thus it has the proper balance of both effects, leading to the highest airfoil efficiency ($C_l/C_d$).

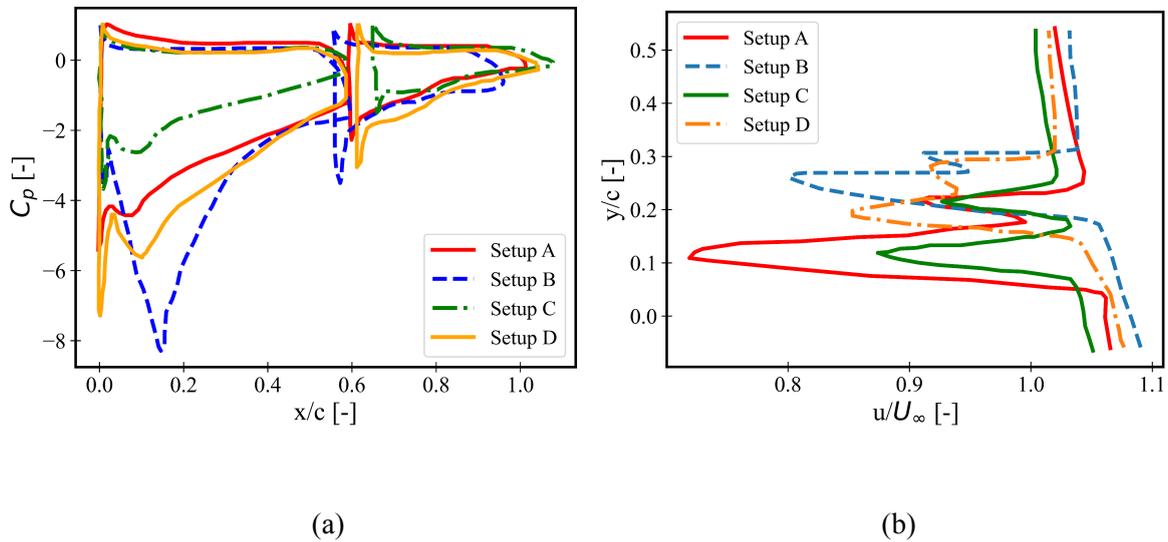

(a)            (b)

Fig 7: (a) Pressure distribution along the surface of the main element and the flap, and (b) Wake velocity profiles at x=2%c downstream of the leading edge of the main element for different setups

## 4  Conclusion

This study presented a multidisciplinary optimization methodology for the design of an inverted double-element airfoil in ground effect for applications in high-performance vehicles. The Improved Harris Hawks Optimization algorithm was employed along with the Kriging surrogate model to obtain the most optimal wing design for three cases. The observations showed that the proposed method led to an improvement of at least 35% in all three cases. The ability of the Improved HHO algorithm and its speed of convergence were further compared with more widely used algorithms such as Particle Swarm Optimization, Genetic Algorithm, Cuckoo Search Algorithm, and standard HHO algorithm.